\documentclass[12pt]{article}
\usepackage{epsfig,graphicx,color}
\usepackage{epstopdf}
\usepackage{latexsym}
\usepackage{amstext}
\usepackage{amsmath,amsfonts,amssymb}
\usepackage[normalem]{ulem}
\usepackage{subfigure}
\usepackage{epsfig, graphicx}
\usepackage{color}
\usepackage{amsfonts,bm,color}
\usepackage{verbatim}
\usepackage{float}
\usepackage{latexsym}
\usepackage{bigints}
\usepackage[english]{babel}
\usepackage{mathrsfs}
\usepackage{bm}
\newcommand{\mb}[1]{ \mbox{\boldmath$#1$} }
\newcommand{\eps}{\varepsilon}
\newcommand{\ds}{\displaystyle}
\newcommand{\beq}{\begin{eqnarray}

x}
\newcommand{\eeq}{\end{eqnarray}}
\newcommand{\beqq}{\begin{eqnarray*}}
\newcommand{\eeqq}{\end{eqnarray*}}

\newcommand{\x}{\mbox{\boldmath$x$}}

\begin{document}

\begin{center}
{\bf{On the spatial coordinate measurement of two identical particles}}\\[2mm]

Avi Marchewka and Er'el Granot\\
Department of Electrical and Electronics Engineering\\
Ariel University, Ariel, Israel\\
avi.marchewka@gmail.com,\ erelgranot@gmail.com\\[3mm]

Zeev Schuss\\
 Department of Mathematics, Tel-Aviv University\\
Tel-Aviv, Ramat-Aviv, 69978, Israel\\
email: schuss@post.tau.ac.il
\end{center}

\vspace{0.5cm}

\begin{abstract}
Theoretically, the coordinate measurement of two identical particles at a point by two narrowly separated narrow detectors, is interpreted in the limit of shrinking width and separation, as the detection of two particles by a single narrow detector. { Ordinarily, the ratio between probabilities of point measurements is independent of the width of the narrow detectors.} We show here that not only this is not the case, but that in some scenarios the results depend on the way the dimensions shrink to zero. The ratio between the width and the separation determines the detection result. { In particular, it is shown that the bunching parameter of bosons is not a well-defined physical property. Moreover, it may suggests that } there is a difficulty in quantum measurement theory in the interpretation of coordinate measurement of two particles.

\end{abstract}


\section{Introduction}\label{sec:intro}

The measurement of the spatial coordinate of a particle is notoriously problematic in quantum theory. Thus,
spatial coordinate measurement reduces the wave function to one of the spatial eigenstates, which is
assumed to be a delta function; however, a delta function is not square integrable and cannot be
normalized, and therefore has no physical meaning. Nevertheless, this problem is avoided by realizing
that pure spatial coordinate measurement is not physical either and, in fact, any detector must have finite
dimensions. Consequently, the delta functions can be approximated by narrow step functions of finite width
\cite{coll}. This choice of approximate eigenstates resolves the problem of the
definition of coordinate measurement of a single particle at a point for any practical purposes.
Nevertheless, spatial coordinate measurements may lead to divergence in some measurable quantities. The
issue of spatial coordinate measurement of a single particle was discussed elsewhere \cite{wigner},
\cite{Hegerfeldt1}, \cite{Hegerfeldt2}, \cite{del campo}, \cite{marchewka Schuss}, \cite{granot marchewka},
and is not addressed here.

The device of approximate eigenstates of the spatial coordinate of two identical particles at a point,
however, does not solve the measurement problem. Specifically, evidence for the difficulty in defining coordinate measurement of two identical particles at a point can be traced to \cite{MarchewkaGranot}, where unexpected bosons
anti-bunching and fermions bunching emerge. It is shown below that two different bunching measurements,
which, according to \cite{Feynman} are expected to be equivalent, lead to different results. It is therefore the purpose of this paper to explore the significance of this behavior. We show, among others, that due to
the fact that
two equivalent measurements { ratios} yield different results, { there is an anomaly in} the measurement of the spatial coordinate of two
identical particles at a point. { Moreover, it suggests that the bunching parameter is not a well-defined parameter}.

\section{Traditional approaches and Feynman's derivation}
In an attempt to demonstrate bosons bunching, Feynman has suggested two approaches to local measurement of two particles \cite{Feynman}.
 In the first approach, the two particles are measured by two single-particle detectors, i.e., each one of the detectors detect only a single particle, and since the two detectors are extremely narrow and adjacent to one another, then double measurement correspond to local measurement of the two particles. The second approach was to take a finite width single two-particle detector. He has shown that in both approaches, in the limit where the size of the detectors goes to zero, the approaches are consistent, and both manifest with bosons bunching.
 Mathematically, for the probability densities of bosons, fermions, and distinguishable particles, ($p^{\mbox{\tiny(bos)}}$, $p^{\mbox{\tiny(fer)}}$, and $p^{\mbox{\tiny(dis)}}$, respectively, see definitions in section (\ref{ss:background}) below), Feynman results can be written
 \begin{align}
 p^{\mbox{\tiny(bos)}}=2p^{\mbox{\tiny(dis)}},\label{bunching}
 \end{align}
 for {\emph both} approaches. Traditionally, this result was interpreted as the bosons' property to bunch.
  Similarly, the classical relation for fermions,
 \begin{align}
 p^{\mbox{\tiny(fer)}}=0,\label{antibunching}
 \end{align}
 is interpreted as the exclusion of fermions (anti bunching).

 Therefore, according to these examples, the two measurement approaches yield the same result, and therefore no inconsistencies occur, and it seems that at least in this sense, the local measurement is well-defined.
 { Feynman has also generalized the bunching effect to $N$ particles (bosonic enhancement factor). However, when the number of final states is also larger than 2 the problem's complexity increases considerably (see for example, Refs.\cite{Lim Beige} \cite{Tichy Tiersch}).}
 Moreover, when the initial states are not orthogonal Eq.(\ref{bunching}) changes to $p^{\mbox{\tiny(bos)}}=\beta p^{\mbox{\tiny(dis)}}$, where $1\le\beta<2$ (see Ref.\cite{marchew_granot_non_orthogonal}).

 \section{Indication of a problem}

 Recently it has been demonstrated that, while Feynman's argument is correct in most cases, in some specific ones it fails (for details see \cite{MarchewkaGranot}, \cite{MarchewkaGranot ann}). For example, when one of the wavefunctions of the particles has a zero, then the two approaches lead to different results.
 The first, two detectors approach, leads to the following result:

\begin{align}
p^{\mbox{\tiny(bos)}}=0,\label{b_antibunching}
\end{align}
and
\begin{align}
p^{\mbox{\tiny(fer)}}=2p^{\mbox{\tiny(dis)}},\label{f_bunching}
\end{align}
which can be interpreted as bosons anti-bunching and fermions bunching, which contradicts Feynman's result.

Moreover, when the second approach is chosen, i.e. a single two-particles detector, the following result {appears}:

\begin{align}
p^{\mbox{\tiny(bos)}}=p^{\mbox{\tiny(fer)}}=p^{\mbox{\tiny(dis)}},\label{no_bunching}
\end{align}
in which case no bunching occurs for either bosons or fermions.

Besides the question of bunching (which was discussed in detailed in \cite{MarchewkaGranot}), there is a disturbing problem of inconsistencies between the two approaching methods.
Since the two measuring approaches are equivalent (they both measure the probability to detect two particles at the same local spot) one would expect, that not only that bosons would always bunch, but that there should be no difference in the outcome of both approaches. It seems that if the local measurement is well defined then they should yield the same result.

One may argue that the source of this discrepancy is the fact that there is something fundamentally different between the two approaches. They are based on two different experimental scenarios: in one experiment there are two detectors and in the next there is only one.
Parenthetically, it should be stressed that there is no reason to expect that the result would be different, after all, when the detectors width and the distance between them shrink to zero, the two scenarios measure the same thing - the probability to detect two particles at the same place.It is clear that in the case of single particle detection the two processes are completely identical.

In this paper we show that the root of the discrepancy is even deeper. The discrepancy still holds even between two similar detection scenarios, which both are local and include two detectors.
We focus on the two detectors scenario. This detection scenario is characterized by two parameters: the width of the detectors and the distance between them. In the limit of local measurement both parameters (detectors width and distance) should shrink to zero. In this paper we show, that in certain cases, the result of the local measurement depends on the ratio between the two parameters. {\emph Every ratio yields a different result}, despite the fact that in all cases they both {go} to zero, i.e.,  it seems that the inconsistency is a fundamental problem in local measurement of two particles.

 \section{Theoretical Background} \label{ss:background}
In Quantum Mechanics there is a clear distinction between distinguishable particles, bosons and fermions.

In the case of spatial coordinate measurement the wave function of two distinguishable particles is reduced to the product \cite{coll},
\begin{align}
\psi^{\mbox{\tiny(dis)}}(x_1,x_2)=\psi_1(x_1)\psi_2(x_2) \mbox{ or } \psi_1(x_2)\psi_2(x_1),
\end{align}
where the single particle wave functions (SPWF's) are $\psi_1(x)$ and $\psi_2(x)$. { The validity of this reduction is independent of the SPWF's orthogonality and of the particles' interaction.}
If it is known that only one of the two particles is in state $\psi_1(x)$, but it is not known which one, the joint pdf is
\begin{align}
p^{\mbox{\tiny(dis)}}(x_1,x_2)=\frac12\left(|\psi_1(x_1)|^2|\psi_2(x_2)|^2+
|\psi_2(x_1)|^2|\psi_1(x_2)|^2\right),\label{a}
\end{align}
which can be written as
\begin{align}
p^{\mbox{\tiny(dis)}}(x_1,x_2)=\frac12\left[p_1(x_1)p_2(x_2)+p_1(x_2)p_2(x_1)\right],\label{pdist}
\end{align}
where the single particle probability density function (pdf) is $p_j(x_i)=|\psi_j(x_i)|^2$ for $i,j=1,2.$

If the two particles are indistinguishable, such as bosons in the same spin states, the joint symmetric wave function is
\begin{align}
\psi^{\mbox{\tiny(bos)}}(x_1,x_2)=\frac{1}{\sqrt{2}}\left[\psi_1(x_1)\psi_2(x_2)+\psi_1(x_2)\psi_2(x_1)\right]
\end{align}
and the joint pdf is \cite{coll}
\begin{align}
p^{\mbox{\tiny(bos)}}(x_1,x_2)=&\frac12\left[p_1(x_1)p_2(x_2)+p_1(x_2)p_2(x_1)+2\mathfrak{R}\mbox{e}
\left\{\psi_1(x_1)\psi^{*}_1(x_2)\psi_2(x_1)\psi^{*}_2(x_2)\right\}\right]\nonumber\\
=&p^{\mbox{\tiny(dis)}}(x_1,x_2)+p^{\mbox{\tiny (inter)}}(x_1,x_2).\label{pbos}
\end{align}
Similarly, for fermions we obtain the antisymmetric joint wave function
\begin{align}
\psi^{\mbox{\tiny(fer)}}(x_1,x_2)=\frac{1}{\sqrt{2}}
\left[\psi_1(x_1)\psi_2(x_2)-\psi_1(x_2)\psi_2(x_1)\right]
\end{align}
and
\begin{align}
p^{\mbox{\tiny(fer)}}(x_1,x_2)=&\frac12\left[p_1(x_1)p_2(x_2)+p_1(x_2)p_2(x_1)-2\mathfrak{R}\mbox{e}
\left\{\psi_1(x_1)\psi^{*}_1(x_2)\psi_2(x_1)\psi^{*}_2(x_2)\right\}\right]\nonumber\\
=&p^{\mbox{\tiny(dis)}}(x_1,x_2)-p^{\mbox{\tiny (inter)}}(x_1,x_2).\label{pferm}
\end{align}

It should be stressed that these equations are valid, provided the single particle wave-functions $\psi_1(x)$ and $\psi_2(x)$ are orthogonal. For non-orthogonal initial states the results could be dramatically different (this case will be discussed elsewhere), but are not relevant to the present work.

\subsection{Measurement of the spatial coordinate of two identical particles at a point}\label{ss:localMeasure}

If the single particle wave function $\psi(x)$ is continuous in the real segment $S: x_a<x<x_b$, then, as $x_a$ and $x_b$ converge to the point $x_0$, { the mean probability density of detecting the particle within \emph{this segment}},
\begin{equation}
\frac{\ds\int_{S}|\psi(x)|^2\,dx}{\int_{S}\,dx}
\end{equation}
becomes independent of { the width of} $S$ { (namely, independent of $x_b-x_a$)}, provided it is sufficiently narrow.
This implies that for two segments, $S_1$ and $S_2$ containing the point $(\xi_1,\xi_2)$,
if $\psi(x_1,x_2)$ is the two-dimensional continuous two-particle wave function (either fermions, bosons or distinguishable particles), then
 \begin{equation}
\frac{\ds\int_{S_1\times S_2}|\psi(x_1,x_2)|^2\/\,dx_1dx_2}{\ds\int_{S_1}\,dx_1\int_{S_2}\,dx_2}\label{local_measur_def}
\end{equation}
becomes independent of $S_1$ and $S_2$ as the endpoints of the segments converge to $(\xi_1,\xi_2)$.

Because the bunching parameter is the ratio between the probability to measure two bosons and the
probability to measure two distinguishable particles at the same point $(\xi_1,\xi_2)$, this ratio is meaningful only if
\begin{equation}
\frac{\ds\int_{S_1\times S_2}|\psi^{(bos)}(x_1,x_2)|^2\,dx_1dx_2}{\int_{S_1\times S_2}|\psi^{(dis)}(x_1,x_2)|^2\,dx_1dx_2} \label{bunching_def}
\end{equation}
is independent of $S_1$ and $S_2$ as the endpoints of the segments converge to $(\xi_1,\xi_2)$.
It is shown below that \eqref{local_measur_def} and \eqref{bunching_def}  are not always valid.

\subsection{Feynman's derivation of boson bunching and fermion anti bunching}\label{ss:classical}

The probability to measure two bosons and two fermions are described by the joint pdf of the coordinates of two identical particles. 
Feynman \cite{Feynman} proposed the following argument, in analogy with the discrete case (see Ref.\cite{Feynman}). Consider the 
single particle wave functions near $x_0$ such that
\begin{align}
\psi_j(x)=&\,\psi_j(x_0)+O(|x-x_0|)\hspace{0.5em}\mbox{with}\ \psi_j(x_0)\neq0 for j=1,2,\label{neq0}
 \end{align}
{ which means, as usual, that there exist $\delta>0$ and $\gamma>0$ such that if $0<|x-x_0|<\delta$, then  
\begin{align}
\left|\frac{\psi_j(x)-\psi_j(x_0)}{x-x_0}\right|\leq \gamma.
\end{align} 
{ It follows, in particular, that if $0<|x_j- x_0|<\delta$ for $j=1,2$, then}
\begin{align*}
&\,p_j(x_j)=|\psi_j(x_0)|^2+O(|x_j-x_0|)\hspace{0.5em} \\
&\,p^{\mbox{\tiny (inter)}}=|\psi_1(x_0)|^2|\psi_2(x_0)|^2
+O(|x_1-x_0|+|x_2-x_0|).
\end{align*}

An alternative approach to the problem of measuring the two particles was suggested in \cite{Feynman} (see Fig.1). It consists in 
detecting them with two detectors of width $\Delta\ll\lambda$, placed at distance $\eta\ll \lambda$ apart, where the particular case 
of two point detectors was considered, that is, $\Delta\ll\eta\ll \lambda$. {For the case at hand $\lambda$ is the smallest
length scale of the wave function. For a plane wave, $\lambda$ is the wave length}.  It has been shown in \cite{MarchewkaGranot} that 
this approach in the scenario given in the previous section can lead to bosons anti bunching and fermions bunching. It turns out that 
the limits of the probabilities for $\Delta\to0$ and $\eta\to0$ are not interchangeable.

Indeed, choosing the condition \eqref{neq0} for $|x-x_0|<\Delta\ll1$, as in \cite{Feynman}, we obtain from
\eqref{pbos}  that for fixed $\eta>0$
\begin{align}
\frac{P^{\mbox{\tiny(bos)}}}{P^{\mbox{\tiny(dis)}}}=\frac{\ds\int_{|x_1-x_0+\eta|<\Delta}dx_1
	\int_{|x_2-x_0-\eta|<\Delta}dx_2\,p^{\mbox{\tiny(bos)}}(x_1,x_2)}{\ds\int_{|x_1-x_0+\eta|<\Delta}dx_1
	\int_{|x_2-x_0-\eta|<\Delta}dx_2\,p^{\mbox{\tiny(dis)}}(x_1,x_2)}=&\,2+O(\Delta^3)\label{bos:dist_2det}\\
\frac{P^{\mbox{\tiny(fer)}}}{P^{\mbox{\tiny(dis)}}}=\frac{\ds\int_{|x_1-x_0+\eta|<\Delta}dx_1
	\int_{|x_2-x_0-\eta|<\Delta}dx_2\,p^{\mbox{\tiny(fer)}}(x_1,x_2)}{\ds\int_{|x_1-x_0+\eta|<\Delta}dx_1
	\int_{|x_2-x_0-\eta|<\Delta}dx_2\,p^{\mbox{\tiny(dis)}}(x_1,x_2)}=&\,O(\Delta).\label{ferm:dist_2det}
\end{align}

Therefore, it is clear that in the ordinary scenarios, which were presented by Feynman, both detecting approaches yield the same result. This fact apparently suggests that the local measurement of a pair of particles is well defined. However, as can be seen from the following section, in some specific cases the two approaches can be inconsistent.

\begin{figure}
	\centering
	\includegraphics[width=8cm,angle=0]{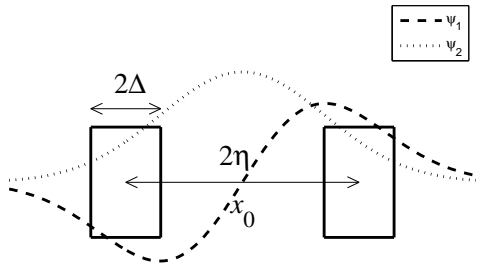}
	\caption{\small The two rectangles represent the detectors and $\psi_1$ and $\psi_2$ are the two wave functions. The distance between the centers of the is $2\eta$ and their width is $2\Delta$. In the case of spatial coordinate measurement measurement, these both distances, $\eta$ and  $\Delta$, shrink to zero.}
\end{figure}

\subsection{A counterexample}\label{ss:counter}
Obviously, \eqref{neq0} does not exhaust all the possibilities and different assumptions lead to different results. The case considered below provides a counterexamples to \eqref{bunching} and \eqref{antibunching} (see \cite{MarchewkaGranot}). Consider the case
\begin{align}
\psi_1(x)=&\,\psi_1'(x_0)(x-x_0)+O((x-x_0)^2)\hspace{0.5em}\mbox{with}\ \psi_1'(x_0)\neq0\hspace{0.5em}\mbox{for}\ x\to x_0 \nonumber\\
\psi_2(x)=&\,\psi_2(x_0)+O(|x-x_0|)\hspace{0.5em}\mbox{with}\ \psi_2(x_0)\neq0\hspace{0.5em}\mbox{for}\ x\to x_0.\label{eq0}
\end{align}
Then
\begin{align*}
&\,p_1(x_1)=|\psi_1'(x_0)|^2(x_1-x_0)^2+O((x_1-x_0)^3)\\
&\,p^{\mbox{\tiny (inter)}}=O((x_1-x_0)(x_2-x_0)),
\end{align*}

It is concluded in \cite{Feynman} is that these results confirm that bosons bunch and fermions anti bunch. It is easy to see, however, that under the assumption \eqref{eq0},
\begin{align}
\frac{P^{\mbox{\tiny(bos)}}}{P^{\mbox{\tiny(dis)}}}=\frac{\ds\int_{|x_1-x_0+\eta|<\Delta}dx_1
	\int_{|x_2-x_0-\eta|<\Delta}dx_2\,p^{\mbox{\tiny(bos)}}(x_1,x_2)}{\ds\int_{|x_1-x_0+\eta|<\Delta}dx_1
	\int_{|x_2-x_0-\eta|<\Delta}dx_2\,p^{\mbox{\tiny(dis)}}(x_1,x_2)}=&\,\frac{\Delta^2}{3\eta^2+\Delta^2}\label{bos:dist_2det2}\\
\frac{P^{\mbox{\tiny(fer)}}}{P^{\mbox{\tiny(dis)}}}=\frac{\ds\int_{|x_1-x_0+\eta|<\Delta}dx_1
	\int_{|x_2-x_0-\eta|<\Delta}dx_2\,p^{\mbox{\tiny(fer)}}(x_1,x_2)}{\ds\int_{|x_1-x_0+\eta|<\Delta}dx_1
	\int_{|x_2-x_0-\eta|<\Delta}dx_2\,p^{\mbox{\tiny(dis)}}(x_1,x_2)}=&\,2-\frac{\Delta^2}{3\eta^2+\Delta^2}.\label{ferm:dist_2det2}
\end{align}
Clearly, changing the order of the limits changes the final result. In particular,
\begin{align}
\lim_{\eta \rightarrow 0}\frac{P^{\mbox{\tiny(bos)}}}{P^{\mbox{\tiny(dis)}}}=1 \label{bos_like_dist}\\
\lim_{\eta \rightarrow 0}\frac{P^{\mbox{\tiny(fer)}}}{P^{\mbox{\tiny(dis)}}}=1, \label{fer_like_dist}
\end{align}
which corresponds to the case of a single detector, as discussed in the previous section, and therefore there is no difference in the properties of measuring bosons, fermions or distinguishable particles in this case.

In the reverse limit,
\begin{align}
\lim_{\Delta \rightarrow 0}\frac{P^{\mbox{\tiny(bos)}}}{P^{\mbox{\tiny(dis)}}}=0 \label{bos_anti_bunch}\\
\lim_{\Delta \rightarrow 0}\frac{P^{\mbox{\tiny(fer)}}}{P^{\mbox{\tiny(dis)}}}=2, \label{fer_bunch}
\end{align}
which means that fermions anti bunch and bosons bunch \cite{MarchewkaGranot}.
Thus
\begin{align}
1=\lim_{\Delta \rightarrow 0}\lim_{\eta \rightarrow 0}\frac{P^{\mbox{\tiny(bos)}}}{P^{\mbox{\tiny(dis)}}} \neq
\lim_{\eta \rightarrow 0}\lim_{\Delta \rightarrow 0}\frac{P^{\mbox{\tiny(bos)}}}{P^{\mbox{\tiny(dis)}}}=0
\label{bos_not_equal1}
\end{align}
and
\begin{align}
1=\lim_{\Delta \rightarrow 0}\lim_{\eta \rightarrow 0}\frac{P^{\mbox{\tiny(fer)}}}{P^{\mbox{\tiny(dis)}}} \neq\lim_{\eta \rightarrow 0}\lim_{\Delta \rightarrow 0}\frac{P^{\mbox{\tiny(fer)}}}{P^{\mbox{\tiny(dis)}}}=2.
\label{bos_not_equal2}
\end{align}
In the distinguished limit $\eta=a\Delta\to0$,
\begin{align}
\frac{P^{\mbox{\tiny(bos)}}}{P^{\mbox{\tiny(dis)}}}=&\,\frac{1}{1+3a^2} \label{bos_siuml}\\
\frac{P^{\mbox{\tiny(fer)}}}{P^{\mbox{\tiny(dis)}}}=&\,2-\frac{1}{1+3a^2}. \label{fer_simul}
\end{align}

Note first, that in a neighborhood of a point that is not a zero of the single particle wave function, as described above, the equations analogous to \eqref{bos_not_equal1} and \eqref{bos_not_equal2} are, respectively,
\begin{align}
\lim_{\Delta \rightarrow 0}\lim_{\eta \rightarrow 0}\frac{P^{\mbox{\tiny(bos)}}}{P^{\mbox{\tiny(dis)}}} =
\lim_{\eta \rightarrow 0}\lim_{\Delta \rightarrow 0}\frac{P^{\mbox{\tiny(bos)}}}{P^{\mbox{\tiny(dis)}}}=2
\label{bos_equal1}
\end{align}
and
\begin{align}
\lim_{\Delta \rightarrow 0}\lim_{\eta \rightarrow 0}\frac{P^{\mbox{\tiny(fer)}}}{P^{\mbox{\tiny(dis)}}} =\lim_{\eta \rightarrow 0}\lim_{\Delta \rightarrow 0}\frac{P^{\mbox{\tiny(fer)}}}{P^{\mbox{\tiny(dis)}}}=0.
\label{bos_equal2}
\end{align}





To interpret the different orderings of the limits, we note that taking the limit $\eta\to0$ first corresponds to putting together two detectors of width $\Delta$, thus rendering them a single detector. Taking the limit $\Delta\to0$ first corresponds to measuring the correlation between two detectors $\eta$ apart.

However, \eqref{bos_siuml} and \eqref{fer_simul} suggest that if $1\le a$, then there are always two detectors and still, the final results depend on the value of $a$. This is clearly seen in Fig.2. The ratio between the same measurement of bosons and distinguishable particles can vary dramatically as a function of $a$. These results clearly contradict \eqref{bunching_def}.

The above considerations indicate that there is a problem in the definition of the spatial coordinate  measurement of two identical particles, because apparently equivalent measurement scenarios give different results.
\begin{figure}
	\centering
	\includegraphics[width=8cm,angle=0]{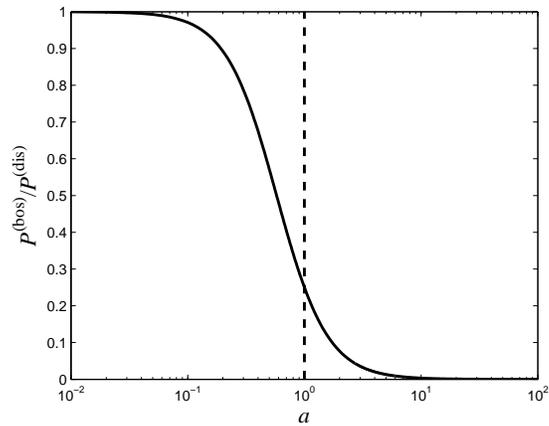}
	\caption{\small The ratio between the probability to measure the spatial coordinate of two bosons at a point and the probability to measure the same for two distinguishable particles as a function of the parameter $a$ (Eq.\ref{bos_siuml}). Clearly, this measurable ratio does not have a single value, but in fact can have any value between 0 and 1. For $a>1$ there is no overlap between the detectors and still the ratio of the local measurements depends on $a$. \label{figure_plot}}.
\end{figure}

A similar anomaly occurs when one calculates the ratio between the probabilities to measure simultaneously one boson by the left detector and one by the right one ($P^{\mbox{\tiny(bos)}}_{\mbox{\tiny{1 left and 1 right}}}$), and the probabilities to measure both of them by the same detector (either left or right) ($P^{\mbox{\tiny(bos)}}_{\mbox{\tiny{2 left or 2 right}}}$). For distinguishable particles this ratio is 1, regardless of the measuring point, however,
\begin{align*}
	\frac{P^{\mbox{\tiny(bos)}}_{\mbox{\tiny{1 left and 1 right}}}}{P^{\mbox{\tiny(bos)}}_{\mbox{\tiny{2 left or 2 right}}}}=\frac{\ds\int\limits_{|x_1-x_0+\eta|<\Delta}dx_1
		\int\limits_{|x_2-x_0-\eta|<\Delta}dx_2\,p^{\mbox{\tiny(bos)}}(x_1,x_2)}{\ds\int\limits_{|x_1-x_0+\eta|<\Delta}dx_1
	\int\limits_{|x_2-x_0+\eta|<\Delta}dx_2\,p^{\mbox{\tiny(bos)}}(x_1,x_2)}=&\,\frac{\Delta^2}{6\eta^2+\Delta^2}\\
	\frac{P^{\mbox{\tiny(fer)}}_{\mbox{\tiny{1 left and 1 right}}}}{P^{\mbox{\tiny(fer)}}_{\mbox{\tiny{2 left or 2 right}}}}=\frac{\ds\int\limits_{|x_1-x_0+\eta|<\Delta}dx_1
		\int\limits_{|x_2-x_0-\eta|<\Delta}dx_2\,p^{\mbox{\tiny(fer)}}(x_1,x_2)}{\ds\int\limits_{|x_1-x_0+\eta|<\Delta}dx_1
\int\limits_{|x_2-x_0+\eta|<\Delta}dx_2\,p^{\mbox{\tiny(fer)}}(x_1,x_2)}=&\,\frac{6\eta^2+\Delta^2}{\Delta^2}.
\end{align*}
We see again that this ratio, which is expected to be 1, depends on the ratio between $\eta$ and $\Delta$. In fact, only when the two detectors overlap completely, that is, $\eta=0$, the ratio is 1.

Again, changing the order of the limits changes the final result. Clearly,
\begin{align*}
\lim_{\eta \rightarrow 0}\frac{P^{\mbox{\tiny(bos)}}_{\mbox{\tiny{1 left and 1 right}}}}{P^{\mbox{\tiny(bos)}}_{\mbox{\tiny{2 left or 2 right}}}}=&\,1\\
\lim_{\eta \rightarrow 0}\frac{P^{\mbox{\tiny(fer)}}_{\mbox{\tiny{1 left and 1 right}}}}{P^{\mbox{\tiny(fer)}}_{\mbox{\tiny{2 left or 2 right}}}}=&\,1,
\end{align*}
as expected. However, in the reverse limit,
\begin{align*}
\lim_{\Delta \rightarrow 0}\frac{P^{\mbox{\tiny(bos)}}_{\mbox{\tiny{1 left and 1 right}}}}{P^{\mbox{\tiny(bos)}}_{\mbox{\tiny{2 left or 2 right}}}}=&\,0\\
\lim_{\Delta \rightarrow 0}\frac{P^{\mbox{\tiny(fer)}}_{\mbox{\tiny{1 left and 1 right}}}}{P^{\mbox{\tiny(fer)}}_{\mbox{\tiny{2 left or 2 right}}}}=&\,\infty. 
\end{align*}
Thus, spatial coordinate measurement when both limits are taken losses its meaning, because
\begin{align*}
1=\lim_{\Delta \rightarrow 0}\lim_{\eta \rightarrow 0}\frac{P^{\mbox{\tiny(bos)}}_{\mbox{\tiny{1 left and 1 right}}}}{P^{\mbox{\tiny(bos)}}_{\mbox{\tiny{2 left or 2 right}}}} \neq
\lim_{\eta \rightarrow 0}\lim_{\Delta \rightarrow 0}\frac{P^{\mbox{\tiny(bos)}}_{\mbox{\tiny{1 left and 1 right}}}}{P^{\mbox{\tiny(bos)}}_{\mbox{\tiny{2 left or 2 right}}}}=0
\end{align*}
and
\begin{align*}
1=\lim_{\Delta \rightarrow 0}\lim_{\eta \rightarrow 0}\frac{P^{\mbox{\tiny(fer)}}_{\mbox{\tiny{1 left and 1 right}}}}{P^{\mbox{\tiny(fer)}}_{\mbox{\tiny{2 left or 2 right}}}} \neq\lim_{\eta \rightarrow 0}\lim_{\Delta \rightarrow 0}\frac{P^{\mbox{\tiny(fer)}}_{\mbox{\tiny{1 left and 1 right}}}}{P^{\mbox{\tiny(fer)}}_{\mbox{\tiny{2 left or 2 right}}}}=\infty.
\end{align*}

In the distinguished limit $\eta=a\Delta\to0$,
\begin{align*}
\frac{P^{\mbox{\tiny(bos)}}_{\mbox{\tiny{1 left and 1 right}}}}{P^{\mbox{\tiny(bos)}}_{\mbox{\tiny{2 left or 2 right}}}}=&\,\frac{1}{1+6a^2}\\
\frac{P^{\mbox{\tiny(fer)}}_{\mbox{\tiny{1 left and 1 right}}}}{P^{\mbox{\tiny(fer)}}_{\mbox{\tiny{2 left or 2 right}}}}=&\,{1+6a^2}. 
\end{align*}
These results { also indicate an anomaly in the simultaneous coordinate measurements of two identical particles, which is not observed in the coordinate measurement of distinguishable particles.} All these limits are consistent with the measurement of the spatial coordinate, however, they all lead to different { ratios of} measurement results.

There are statistical consequences to these considerations. First, because this effect occurs only in a neighborhood of the zeros of 
the single particle wave function, there
should be no spatial averaging, unlike that done in \cite{Aspect} and \cite{18}. Second, the manifestation of the anomaly requires 
that one of the quantum states be locally antisymmetric around a zero. Therefore, the first bound state, which is approximately a 
Gaussian or an approximately spherical wave function, is not a good candidate for the manifestation of this effect, in contrast to 
\cite{19}, \cite{20}, and \cite{21}. It is therefore not surprising that the phenomenon discussed here has not been observed 
experimentally so far.

\section{Discussion, conclusions, and generalization}

The paper reveals discrepancies in the realization of spatial coordinate measurement of two indistinguishable particles (bosons or fermions), due to the wave symmetrization postulate. It demonstrates that when one of the wave functions changes sign at a point, then near this point { there is an anomaly in} the measurement of the spatial coordinate of two identical particles. However, this anomaly does not occur for distinguishable particles. It is known \cite{Feynman} that if two detectors are used, then spatial coordinate measurement requires taking their width as well as the distance between them to zero. However, the analysis presented in this paper shows that the order of the limits and the ratio between these parameters has a large effect on the final result. { In particular, it is shown that the bunching parameter is not a well-defined.}

{
It should be emphasized that despite the fact that we have focused in this paper on the limits of these expressions, the effect can be realized experimentally even for finite dimensions. As long as $\eta<<\lambda$ and $\Delta<<\lambda$ it is sufficient to change the ratio between $\eta$ and $\Delta$ to witness the anomaly.}

Note that assumption \eqref{eq0} is a particular case of the more general assumption that the joint wave function (in higher dimensions as well)  $\psi_1(\x)/\psi_2(\x)$ is antisymmetric in a small ball centered at $\x$ in the sense that
$\psi_1(\x+\eps\mb{e})\psi_2(\x-\eps\mb{e})=-\psi_1(\x-\eps\mb{e})\psi_2(\x+\eps\mb{e})+o(\eps)$ for all vectors $\mb{e}$ such that $|\mb{e}|=1$ and $\eps\ll1$.
Moreover, it should be further noted that the above discussion can be generalized from spatial coordinate, which is important for realizations, to any continuous observable, such as momentum.


\end{document}